\documentclass[11pt]{article}
\usepackage{fullpage}
\usepackage{amsmath}
\usepackage{amssymb}
\usepackage{algorithmic}
\usepackage{mathrsfs}
\usepackage{xcolor}   
\usepackage{microtype}                                                                                                                                                                                
\usepackage{enumerate}
\usepackage{soul}
\usepackage{url}

\newcommand{\algolabel}[1]{\newcounter{#1} \setcounter{#1}{\value{ALC@line}}}
\newcommand{\algoref}[1]{\arabic{#1}}
\newenvironment{Proof}{\noindent{\bf Proof.}}{\hfill$\Box$\FF}
\newenvironment{proofsketch}{\noindent{\bf Proof
sketch.}}{\hfill$\Box$\FF}
\newtheorem{theorem}{Theorem}
\newtheorem{lemma}{Lemma}
 \newtheorem{fact}{Fact}
\newtheorem{definition}{Definition}

\newcommand{\BB}{\vspace*{-\medskipamount}}

\newcommand{\Section}[1]{\section{#1}}

\newcommand{\FF}{\vspace*{\medskipamount}}

\newcommand{\T}{\hspace{1em}}   
\newcommand{\TT}{\hspace*{2em}}
\newcommand{\TTT}{\hspace*{3em}}

\newcommand{\co}[1]{\emph{/* #1 */}}

 \newcommand{\act}[1]{\mbox{#1}}

\newcommand{\tup}[1]{\langle #1 \rangle}
\newcommand{\whp}{\textit{whp}}

\newtheorem{observation}{Observation}
\newcommand{\remove}[1]{}
\newtheorem{assert}{Assertion}[section]

\newcommand{\ENed}{enlightened}

\newcommand{\sglspc}[1]{\renewcommand{\baselinestretch}{#1}\normalsize}
\newcommand{\p}[1]{\vspace{0.45\smallskipamount}\noindent{\bf #1}}

\newcommand{\M}[1]{${\cal F}_{{\it#1}}$}
\newcommand{\Aest}{$A_{\it est}$}
\newcommand{\AEalgo}{\Aest}
\newcommand{\epoch}[1]{epoch~$\!\mathfrak{#1}$}
\newcommand{\SRA}{{\it SRA}}

\begin{document}

\title{Technical Report:
Estimating Reliability of Workers\\ for Cooperative Distributed Computing\thanks{
This work is supported in part by the NSF award 1017232.}
}

\author{
  Seda Davtyan$^*$%
 	 \thanks{$^*$
 	      Department of Computer Science \& Engineering,
         University of Connecticut,
         371 Fairfield Way, Unit 4155,
 	      Storrs CT 06269, USA.
         Emails: {\tt \{seda,aas\}@engr.uconn.edu}.
  } 
 ~ ~ Kishori M. Konwar$^\dag$%
  \thanks{$^\dag$
           University of British Columbia,
           Vancouver, BC V6T 1Z3, CANADA
           Email: {\tt kishori@interchange.ubc.ca}.
 }
 ~ ~ Alexander A. Shvartsman$^*$
  }

\date{}


\thispagestyle{empty}
\maketitle

\begin{abstract}
Internet supercomputing is an approach to solving partitionable,
computation-intensive problems by harnessing the power of
a vast number of interconnected computers.
For the problem of using network supercomputing to perform
a large collection of independent tasks,
prior work introduced a decentralized approach  
and provided randomized synchronous algorithms that
perform all tasks correctly with high probability,
while dealing with misbehaving or crash-prone processors.
The main weaknesses of existing algorithms is that they assume either that
the \emph{average} probability 
of a non-crashed processor returning incorrect results is inferior to $\frac{1}{2}$,
or that the probability of returning incorrect results is known
to \emph{each} processor.
Here we present a randomized synchronous distributed algorithm that tightly estimates the
  probability of each processor returning correct results.
Starting with the set $P$ of $n$ processors, let $F$ be the set of
processors that crash.
Our algorithm estimates  the  probability $p_i$ of returning a correct result
for each processor $i \in P-F$,
making the estimates available to all these processors.
The estimation is based on the $(\epsilon, \delta)$-approximation,
where each estimated probability $\tilde{p_i}$ of $p_i$ obeys the bound
${\sf Pr}[p_i(1-\epsilon) \leq \tilde{p_i} \leq p_i(1+\epsilon)] > 1 - \delta$, 
for any  constants $\delta >0$ and  $\epsilon >0$ chosen by the user.
An important aspect of this algorithm is that each processor 
terminates without global coordination.
We assess the efficiency of the algorithm in three adversarial models as follows.
For the
model where the number of non-crashed processors $|P-F|$ is  linearly bounded
the time complexity $T(n)$ of the algorithm is $\Theta(\log{n})$,
work complexity $W(n)$ is $\Theta(n\log{n})$, and message complexity $M(n)$ is $\Theta(n\log^2n)$.
For the model where $|P-F|$ is bounded
by a fractional polynomial ($|P-F| = \Omega(n^a)$, for a constant $a
\in (0,1)$) we have
$T(n) = O(n^{1-a}\log{n}\log{\log{n}})$,  
$W(n) = O(n\log{n}\log{\log{n}})$, and 
$M(n)=O(n\log^2n)$.
For the model where $|P-F|$ is bounded
by a poly-logarithm we have $T(n)=O(n)$, $W(n)=O(n^{1+a})$, and
     $M(n)=O(n^{1+a})$. 
All bounds are shown to hold with high probability.
\end{abstract}

\section{Introduction}

Cooperative network supercomputing is becoming increasingly popular for harnessing the power
of the global Internet computing platform. A typical Internet supercomputer, e.g., \cite{DISNET,SETIhome}, 
consists of a master computer  and a large number of computers called workers, performing 
computation on behalf of the master. Despite the simplicity and benefits of a single master approach, 
as the scale of such computing environments grows, it becomes unrealistic to assume the existence 
of the infallible master that is able to coordinate the activities of multitudes of workers. 
Large-scale distributed systems are inherently dynamic and are subject to perturbations, such as 
failures of computers and network links, thus it is also necessary to consider fully distributed peer-to-peer solutions.

Interestingly, worker computers returning bogus results is a phenomenon 
of increasing concern. While this may certainly occur unintentionally,
e.g., as a result of over-clocked processors, workers may in fact
deceitfully claim to have performed 
assigned work so as to obtain incentives associated with the system,
e.g.,  a higher rank.
To address this problem,
several works, e.g., 
\cite{NCA2011,FGLS2005b,FGLS2005,KRS2006}, study
approaches based on a reliable master dealing
with a collection of unreliable workers. 
The drawback in these approaches is the reliance  
on a reliable, bandwidth-unlimited master processor.

In our recent work \cite{DKRS2013,DKS2011,DKS2012} we began to address
this drawback of centralized systems by removing the assumption of 
an infallible and powerful master processor.
We introduced a decentralized approach, where
a collection of worker processors cooperates on a large
set of independent tasks without the reliance on
central control. Our synchronous algorithms in~\cite{DKS2011,DKRS2013} are
able to perform all tasks with high probability, while
dealing with misbehaving processors under  
a rather strong assumption that the average probability of live 
(non-crashed) processors returning incorrect results remains 
inferior to $\frac{1}{2}$ during the computation. 
In~\cite{DKS2011} we considered a linearly bounded adversary, 
where the adversary is constrained so that a constant fraction of workers never fails.
Our algorithm in~\cite{DKRS2013}  works additionally for adversaries
constrained not to to fail the number of workers that is bounded by
a fractional polynomial or a poly-logarithm.
The adversary may assign arbitrary constant probabilities
to processors, provided that the processors remaining in
the computation return bogus results with the
average probability inferior to $\frac{1}{2}$. 
Thus in~\cite{DKS2011} and in~\cite{DKRS2013}, the adversary is
severely limited in its ability to crash  
processors that normally return correct results.
To address this limitation, in~\cite{DKS2012} 
we considered a linearly bounded model where the average probability of
non-crashed processors returning bogus results can become 
greater than $1/2$. 
However, the algorithm in~\cite{DKS2012} assumes that every processor
knows 
the probability of returning an incorrect result for \emph{all} processors. 
This is a very strong assumption, thus it is important to develop
decentralized algorithms that can efficiently estimate these probabilities
in the setting of cooperative distributed computation. 

\p{Contributions.} 
For the general setting of network supercomputing
we consider the problem of estimating the probability of
each participating processor performing a task correctly.
The requirement here is that these estimates are 
computed efficiently in a distributed  system of $n$ workers 
\emph{without} centralized control. 
The estimation is done with the help of  ``test tasks,'' i.e., 
tasks whose results are known to a processor that
needs to obtain these estimates.
Each test task can be performed by any worker in constant time.
For the initial set $P$ of $n$ processors, we assume that every 
processor $i \in P$ is given a distinct set of test tasks $TT_i$, 
for which it knows the correct results, and others do not.
Processors communicate via a synchronous fully-connected message-passing system.
We deal with failure models where workers can return
incorrect results and can crash.
In particular, the adversary assigns to each processor $i$
the probability $p_i$ of returning correct results (incorrect
results are returned with probability $1-p_i$).
We present a randomized decentralized algorithm
 that estimates,  for each processor $i\in P$,  
 the probability $\tilde{p_i}$ of returning correct results,
and such estimates are computed by all processors.
The estimates are calculated using the  $(\epsilon, \delta)$-approximation,
  for  $ 0 <\epsilon < 1$ and $\delta > 0$,
that estimates the mean of a random  variable.
For the given $\delta >0$ and $\epsilon > 0$ chosen by the user,
the algorithm obtains estimates $\tilde{p_i}$ that obey the following bound:
$ {\sf Pr}[p_i(1-\epsilon) \leq \tilde{p}_i \leq p_i(1+\epsilon)] > 1 - \delta$.  
We analyze our algorithm and assess its time, work, and message complexities.
  In additional detail our contributions are as follows.

\p{1.}
We formulate the following model of adversity.
Given the initial set of processors $P$, with $|P|=n$,
the adversary assigns arbitrary constant 
positive probability $p_i$ of performing tasks correctly
to each processor  $i \in P$.
Additionally, the adversary can crash a set $F$ of processors,
subject to one of the  three constraints:
$1)$ 
The adversary is constrained by a  linear fraction, 
where $|P-F| \geq h n$, with  $1 < h < 1-f$ and $f \in (0, 1)$.
$2)$
The adversary is constrained by a fractional polynomial, where
$|P-F| = \Omega(n^a)$, for a constant $a \in (0,1)$.
$3)$
The adversary is constrained by a poly-log,
where $|P-F| = \Omega(\log^c n)$, for a constant $c \geq 1$. 
(Constraints $(2)$ and $(3)$ are as in~\cite{DKRS2013}.)

\p{2.}
We present a randomized algorithm for $n$ processors
to compute the estimates of probabilities $p_i$.
The algorithm works in synchronous rounds,
where each processor asks some other
processor to perform a test task and return the result.
It then shares its  knowledge of results 
with \emph{one} randomly chosen processor. 
Once a processor accumulates a ``sufficient" number of results,
it becomes ``enlightened."
Enlightened processors then ``profess" their knowledge by multicasting
it to a random, exponentially growing subsets of processors.
When a processor receives a  
message telling it that ``enough" gossip was done, it halts.
The values that control ``sufficient" numbers of results 
and ``enough" gossiping are established in our analysis and are 
used as \emph{compile-time} constants. 

We consider the protocol, by which the ``enlightened" processors ``profess"
their knowledge and reach termination, to be of independent interest.
The protocol's message complexity does not depend on crashes, and
termination does not require explicit coordination. This addresses
the challenge of termination when $P-F$ can vary
broadly in the considered three models.

\p{3.}
Our analysis shows that in each model  all live processors estimate the
probability $p_i$ for every processor $i \in P-F$ using
the $(\epsilon,\delta)$ approximation, \whp{}
(henceforth we use the notation \whp{} to stand
for ``with high probability").
Complexity results for the algorithm also hold \whp{}:

\begin{itemize}
\item
For the linearly bounded model we show that work
  complexity $W(n)$ is $\Theta(n \log{n})$, message complexity $M(n)$ is $\Theta(n \log^2{n})$,
and time complexity $T(n)$ is $\Theta(\log{n})$.

\item
For the polynomially constrained model we show that 
$W(n)$ = $O(n\log{n}\log{\log{n}})$,
$M(n)$ = $O(n\log^2{n}\log{\log{n}})$,
and $T(n)$ = $O(n^{1-a}\log{n}\log{\log{n}})$. 

\item
For the poly-log constrained model we show that
$W(n)= O(n^{1+a})$,
$M(n)=O(n^{1+a})$,
and $T(n)=O(n)$.
\end{itemize}

The work complexity results show that the algorithm is efficient, e.g.,
if $\Theta(n^{1+a})$ real tasks are to be done after the
estimation, then the estimation expense
is  amortized.

Finally we note that the  $(\epsilon, \delta)$-approximation is rarely seen 
in distributed computing literature, and we consider showing the relevance
of this technique, and bringing it to the attention of researchers in distributed computing,
to be among the contributions of this work.

\p{Prior/Related Work.} 
Earlier approaches explored ways of improving the quality of 
the results obtained from untrusted workers in the settings where
an infallible master is coordinating the workers.
Fernandez et al.~\cite{FGLS2005,FGLS2005b} and 
Konwar et al.~\cite{KRS2006}
present algorithms that help the master determine correct results \whp{},
while minimizing work.
Additionally,~\cite{KRS2006} provides efficient algorithms that 
 can estimate the probability of processors returning incorrect results. However, 
they assume that this probability is the same for every processor.
The failure models assume that some fraction
of processors can exhibit faulty behavior.
Another recent work by Christoforou et al.~\cite{NCA2011} pursues
a game-theoretic approach.
Fernandez et al.~\cite{FGLS2011a} studied the master-worker model with message loss and delays
 in addition to assuming that processors can return incorrect results;
they give algorithms with exact bounds on work and expected work.
Paquette and Pelc~\cite{PP2004} consider a
fault-prone system in which a decision has to be made on the basis of 
unreliable information, and design a deterministic  strategy 
that leads to a correct decision \whp{}.

As already mentioned, 
our prior work \cite{DKRS2013, DKS2011} introduced the decentralized approach that
eliminates the master, and provided
a synchronous algorithm that is able to perform all tasks
\whp{}, while dealing with incorrect behaviors under 
a very strong assumption that the average probability of 
non-crashed processors returning incorrect results remains 
inferior to~$\frac{1}{2}$. 

The  $(\epsilon, \delta)$-approximation
has been applied to a wide range of 
 difficult scientific problems. For example, it has been successfully applied
 for approximation of probabilistic inference in Bayesian networks~\cite{DL1997},
 solving Ising model problems in statistical mechanics~\cite{JS1993},  estimation of 
  convex bodies~\cite{DFK1991}, and estimating the number of solutions to a DNF formula~\cite{KLM1989}. 
We refer the reader to~\cite{DKLR1995} for a broader list of references.

\p{Document structure.}   
In Section~\ref{model} we present the model and measures of efficiency. 
Our algorithm is given in Section~\ref{algorithm}.
In Section~\ref{estimation} we discuss the estimation techniques. 
In Section~\ref{analysis} we analyze the algorithm and derive  
complexity bounds.
We conclude in Section~\ref{conclusion} with a discussion.


\Section{Model of Computation and Definitions} \label{model}

\p{System model.}
There are $n$ processors, each with a unique 
identifier (id) from  set  $P = [n]$. We refer to the processor 
with id $i$ as processor $i$.
The system is synchronous and processors communicate by exchanging 
reliable messages.
Computation is structured in terms of synchronous \emph{steps},
where in each step a processor can send messages, receive messages,
and/or perform local polynomial computation, where the local
computation time is assumed to be negligible compared to message latency. 
Messages received by a processor in a given step include all 
messages sent to it in the previous step. 

\p{Tasks.}
Ultimately the cooperating processors must perform \emph{tasks}.
Each task can be performed locally by any processor.
The tasks are (a) similar, meaning that any task can be done in 
constant time, (b) independent, meaning that each task 
can be performed independently of other tasks, and (c) idempotent, meaning 
that the tasks admit at-least-once semantics and can be performed concurrently. 
To avoid misrepresentation of results, we assume that once a processor performs a task, it
unforgeably signs the result (this is not discussed further).
Lastly, every processor $i \in P$ is given a distinct set of \emph{test tasks} $TT_i$,
for which only it knows the correct results.

\p{Models of adversity.}
Processors are undependable: a processor may
compute  results of tasks incorrectly and it may crash.
Following a crash, a processor
performs no further actions.
Otherwise, each processor adheres to the protocol
of the algorithm it executes.
Messages can be sent to crashed processors, but they
are neither delivered nor a crashed processor responds.
Thus a crash can be detected if an expected response
does not arrive.
We refer to non-crashed processors as \emph{live}.

We consider an oblivious adversary that, prior to the computation, 
$(a)$~assigns an arbitrary constant probability $p_i > 0$ of returning a correct
result for each processor $i \in P$, and
$(b)$~decides what processors to crash and
when to crash them. 
For an execution of an algorithm, let
$F$ be the set of processors that adversary crashes;
the  number of processors
that can crash is established by the following adversarial models.

\noindent
\emph{Model} \M{\ell f}: The adversary is constrained by a
  \emph{fraction} of the processors in $P$:
{
$|P-F| \geq hn$, where $1 < h < 1 -f$ and $f\in (0,1)$, such that, up to $f|P|$
processors can be crashed.
}

\noindent
\emph{Model} \M{fp}: The adversary is constrained by a
  \emph{fractional polynomial}:
{
$|P-F| = \Omega(n^a)$, for a constant $a\in (0,1)$.
}

\noindent
\emph{Model} \M{pl}: The adversary is constrained by a \emph{poly-logarithm}:
{
$|P-F| = \Omega(log^c n)$, for a constant  $c \ge 1$. 
}

\p{Measures of efficiency.}
We assess the efficiency of algorithms in terms of \emph{time} $T(n)$, \emph{work} $W(n)$,
and \emph{message} $M(n)$ complexities.
We use the conventional measures of  {time complexity}, assessed as the maximum number of steps executed by any processor,
and {work complexity}, assessed as the total number of steps executed by all $n$ processors.
We assess  {message complexity} as the number of point-to-point messages 
sent during the execution.
Lastly, we use the common definition of {\em an event} $\mathcal{E}$ 
{\em occurring with high probability} (\whp{}) to mean that
${\sf Pr}[\mathcal{E}] = 1 - O(n^{-\alpha})$ for some constant $\alpha > 0$.


\Section{Algorithm Description} \label{algorithm}

We now present our decentralized algorithm $A_{est}$ that employs
 no master and instead uses a gossip-based approach to share information. 
The algorithm is structured in terms of the main loop that iterates through three stages: 
{\sc query}, {\sc response}, and {\sc gossip}.
Each stage consists of three steps,  
{\em Send}, {\em Receive}, and {\em Compute}, that
are executed synchronously by the processors.
In the {\sc query} stage each processor sends, receives, and performs test tasks. 
During the {\sc response} stage the processor replies with the results for 
the test tasks, if any, and collects such results sent by other processors.
If enough information is gathered, the processor becomes ``enlightened."
In the {\sc gossip} stage each processor gossips the collected results
to one other processor, except that enlightened processors ``profess" their results 
to an exponentially growing random sets of processors.
The processors then update their local knowledge based on the received messages, 
and, if sufficient information was propagated,
compute the estimates for the probabilities $p_i$ and halt.
The pseudocode for  algorithm~\AEalgo{} is given in
Figure~\ref{fig:alg}; the algorithm uses subroutine Estimation() to compute
the probabilities, given in Figure~\ref{fig:subalg}.
We next describe the algorithm in greater detail.

\begin{figure*}[ht!]
\sglspc{0.9}
\hrule \smallskip
\begin{center}
\parbox{0.9\textwidth}{
\begin{algorithmic}[1]
{
\scriptsize
\STATENO{\bf procedure} for  processor $i$;
\STATENO\T {\bf input} $n,$  \co{$n$ is number of  processors}
\STATENO\TT\TT\T $\epsilon, \delta$, \co{$\epsilon >0$ and $\delta >0$ are estimation parameters}
\STATENO\TT\TT\T $TT_i$ \co{the set of test tasks for $i$}
\STATENO\T {\bf output} $Estimate_i[1..n]$ {\bf init}  $\bot$ \co{array of estimates of $p_j$ for each $j\in P$}
\STATENO\T $R_i[1..n]$ {\bf init} $\emptyset^n$ \co{set of collected result indicators $\tup{res,src,rnd}$}
\STATENO\T {\bf int} $r$  {\bf init} $0$ \co{round number}
\STATENO\T {\bf int} $\ell$  {\bf init} $0$ \co{specifies the number of ${\sf profess}$
  messages to be sent per iteration}
\STATENO\T {\bf bool} $enlightened$  {\bf init} ${\sf false}$ \co{indicates whether the processor is ``enlightened"}
\smallskip
\STATENO\T {\bf while}  ${\sf true}$ {\bf do} 

\STATENO\T\T {\sc query stage}
\STATENO\TT\T {\em Send:}
\STATEIND\TTT\T  Let $q$ be a randomly selected processor id from $ {P} $ \algolabel{proc:select}
\STATEIND\TTT\T Let $t$ be a randomly selected task from $TT_i$
\STATEIND\TTT\T  Send $\langle  t, i\rangle$ to processor $q$ \algolabel{send:task} 

\STATENO\TT\T {\em Receive:}
\STATEIND\TTT\T Let $M = \{m : m = \tup{task,id}\}$ be the set of received messages 

\STATENO\TT\T {\em Compute:}
\STATEIND\TTT\T {\bf if} $|M| > \lceil \log{n} \rceil$ {\bf then}
\STATEIND\TTT\T\TT ${M} \gets \act{random selection of}~\lceil \log{n} \rceil~\act{elements from}~M$ 
\STATEIND\TTT\T Let $V=\{\tup{val,id} : m\in M \wedge val = {\rm ~result~of~} m.task \wedge id=m.id\}$ 

\smallskip
\STATENO\T\T {\sc response stage}
\STATENO\TT\T {\em Send:}
\STATEIND\TTT\T {\bf for each} $w \in V$ {\bf do}
\STATEIND\TTT\T\TT Send $\langle w.val \rangle$ to $w.id$

\STATENO\TT\T {\em Receive:}
\STATEIND\TTT\T   {\bf if} message $\tup{val}$ is received from $q$ chosen in {\sc query stage} {\bf then} 
\STATEIND\TTT\T\TT {\bf if}  $val$ is the correct result for task $t$ chosen in {\sc query stage} {\bf then}
\STATEIND\TTT\T\TT\TT $R_i[q] \gets R_i[q] \cup \tup{1,i,r}$ \co{test task was computed correctly}
\STATEIND\TTT\T\TT {\bf else} 
\STATEIND\TTT\T\TT\TT $R_i[q] \gets R_i[q] \cup \tup{0,i,r}\rangle$ \co{test task was computed incorrectly}
\STATEIND\TTT\T {\bf else} \co{no response from processor $q$}
\STATEIND\TTT\T\TT $R_i[q] \gets R_i[q] \cup \tup{-1,i,r}$ \co{ $-$$1$ is used to record a crash}\algolabel{crash}
\STATENO\TT\T {\em Compute:}
\STATEIND\TTT\T  {\bf if} $\forall j\in P : 
(\sum_{x \in R_i[j]}I_{\{1\}}(x.res) \geq \Gamma_1)$ \co{sufficient
  no. of correct results} \algolabel{alg:enl-start}
\STATEIND\TTT\TTT\TT\TT
$\vee (\exists x\in R_i[j] : x.res=-1$) {\bf then} \co{or $j$ crashed}
\STATEIND\TTT\T\T $enlightened \gets {\sf true}$  \co{processor becomes \ENed{}} \algolabel{alg:enl-end}

\smallskip
\STATENO\T\T {\sc gossip stage}
\STATENO\TT\T {\em Send:}
\STATEIND\TTT\T {\bf if} $enlightened$ {\bf then} \co{gossip aggressively}
\STATEIND\TTT\T\T  Let $D$ be a set of $2^{\ell-1} \log{n}$ processor ids randomly selected from
${P} $ \algolabel{alg:proc}
\STATEIND\TTT\T\T Send $\langle {\sf profess}, R_i[\; ], \ell, i
\rangle$ to processors in $D$
\STATEIND\TTT\T\T $\ell \gets \ell + 1$
\STATEIND\TTT\T {\bf else} 
\STATEIND\TTT\T\T  Let $q$ be a randomly selected processor id from $P$
\STATEIND\TTT\T\T Send $\tup{{\sf share}, R_i[\; ], \ell, i}$ to
processor $q$ \algolabel{share}

\STATENO\TT\T {\em Receive:}
\STATEIND\TTT\T Let $M = \{m : m=\tup{type,R,\ell,id}\}$ be the set of received messages
\STATEIND\TTT\T {\bf if} $\exists m \in M$ : $m.type = {\sf profess }$ {\bf then} 
\STATEIND\TTT\T\T $enlightened \gets {\sf true}$  \co{processor becomes \ENed{}} 
\STATEIND\TTT\T {\bf if} $\exists m \in M$ : $(\ell, i)  \prec (m.\ell, m.id)$  {\bf then} \algolabel{alg:lexi}
\STATEIND\TTT\T\T\T $\ell \leftarrow 0$

\STATENO\TT\T {\em Compute:}
\STATEIND\TTT\T  {\bf for each} $j\in P$ {\bf do}
\STATEIND\TTT\T\T $R_i[j] \leftarrow  R_i[j] \cup  \bigcup_{m \in M} m.R[j]$
\STATEIND\TTT\T {\bf if}  $\exists m \in M : m.\ell \geq \lceil \log n \rceil$ {\bf then} \algolabel{alg:stop}
\STATEIND\TTT\T\T  Estimation($R_i[~],Estimate_i[~]$) \co{Compute the estimates and store in $Estimate_i[1..n]$}
\STATEIND\TTT\T\T {\bf halt}
\STATEIND\TTT\T $r \gets r+1$
}

\end{algorithmic}
}
\end{center}
\hrule
\caption{\rm Algorithm $A_{est}$ at processor $i$ for $i \in P$.}\BB\BB
\label{fig:alg}
\end{figure*}

\p{Inputs.}
Each processor $i$ receives as inputs the number of processors $n$, the estimation parameters
$\epsilon$ and $\delta$, and the set of test tasks $TT_i$ from its environment.

\p{Output.}
Each processor $i$ outputs the estimates of probabilities $p_j$ for each $j\in P$
in array $Estimate_i[1..n]$. If a crash of  processor $j$ is detected, 
   $Estimate_i[j]$ is set to $-1$.

\p{Local knowledge and state variables.}
Every processor $i$ maintains the following:

\begin{itemize}
	\item Array $R_i[1..n]$ stores results of test tasks, where 
       element $R_i[j]$  is a set of
      results of test tasks done by processor $j$.  Each $R_i[j]$ is a set of
       tuples $\tup{v,s,r}$ representing the correctness of the result  
	 $v$ ($v \in \{0, 1, -1\}$)  computed by processor $j$ on behalf of processor $s$, in
         round~$r$. 
      (This ensures that results computed by processor $j$ in
      different rounds $r$  and for different processors $s$ are included.) 
       The value $v=0$ means that the result
      was computed incorrectly, $v=1$ means that it was
      computed correctly, and $v=-1$ means that processor $j$
      has not returned a result, hence, per our model assumption, it crashed.

          \item $r$ is the round (iteration) number that is
            used  to timestamp the computed results.

          \item $\ell$ controls the number of messages multicast
            by  enlightened processors: the multicast is sent to $2^{\ell-1}$ destinations.
           The value of $\ell$ is also used to ``prioritize" processors, where higher values of $\ell$ 
correspond to higher priority, with ties broken by the processor identifiers.
That is, given two distinct processors $i$ and $j$ we say that processor $j$ has \emph{higher priority} than 
$i$ if  $(\ell_i, i) \prec (\ell_j, j)$, where $\prec$ is a lexicographic comparison. i.e.,
 $(\ell_i, i) \prec (\ell_j, j)$  if and only if either 
$(i)$ $\ell_i  < \ell_j$,  or 
$(ii)$  $\ell_i = \ell_j$ and  $i < j$.

          \item  {\it enlightened} is a boolean that determines whether the processor has enough information to start
``professing" its knowledge by means of aggressive gossip.
\end{itemize}

\p{Control flow.}
We refer to each iteration of the main while-loop as the \emph{round}.
The loop is synchronous, but each processor exits the loop based on its local state,
thus  the loop may not terminate simultaneously; to model this we let the loop iterate
forever and include an explicit {\bf halt} for each processor~$i$.
Next we detail each of the three stages within a round.
Recall that each stage is comprised of three steps.

\noindent{\sc Query} stage: 

{\emph{Send step:}}
Processor $i \in P$ selects at random a target processor $q \in P$ 
and a task $t \in TT_i$ and sends the request containing task $t$ to $q$.

{\emph{Receive step:}}
The processor receives the requested tasks
sent to it in the preceding step (if any).

{\emph{Compute step:}}
If the number of tasks requested is less than $\lceil \log{n}\rceil$, 
the  processor computes all the tasks received. Otherwise, it randomly 
selects $\lceil \log{n}\rceil$ tasks and computes the results for the selected tasks.
 The results are stored in a temporary set variable $V$ where each element
 is a pair  $\tup{val,id}$, where $val$ is the 
result of the task computed by processor $i$ as requested by processor $id$.
(We will show in the analysis of the algorithm that 
although the algorithm  performs  at most $\lceil \log{n} \rceil$ tasks,
this is sufficient for our estimation \emph{whp}.)

\smallskip
\noindent{\sc Response} stage:

 {\emph{Send step:}}
Based on data in $V$, processor $i$  sends results of tasks to the respective requesters. 

{\emph{Receive step:}}
Processor $i$ receives the result in a message $m$ (if any) from  processor $q$
that it selected in the {\sc query} stage. 
If the result for the test task is correct
then processor $i$ adds $\langle 1,i,r\rangle$ to  $R_i[q]$, otherwise it adds
$\langle 0,i,r\rangle$. If, however, it does not receive a message 
from $q$ then it adds $\langle -1,i,r\rangle$ to $R_i[q]$, where $-1$ 
indicates that processor $q$ crashed.

{\emph{Compute step:}
Processor $i$ uses the values in $R_i[~]$ to check whether 
it gathered a certain number of results
(in the analysis we will show that this is sufficient for computing
the $(\epsilon, \delta)$-approximation).
If so, the processor becomes \ENed{}.
This is done with the help of the function call $I_{\{1\}}(x.res)$ in line~\algoref{alg:enl-start}.
The function $I_A: \mathbb{N} \rightarrow \{0,1\}$ is the \emph{indicator function}, such that $I_A(x)$, for any set 
$A\subseteq \mathbb{N}$,  returns value $1$ if $x \in A$  and $0$ otherwise
(this is also used in the analysis).

\smallskip
\noindent{\sc Gossip} stage:

{\emph{Send step:}}
If processor $i$ is enlightened, it aggressively gossips its knowledge
by professing it to an exponentially growing random set of processors.
The size of the set is governed by the exponent $\ell$ that is incremented
in each round.
Otherwise the processor shares its knowledge with one randomly chosen processor. 

{\emph{Receive step:}}
Processor $i$ receives messages.
If it receives a  ${\sf profess}$ message, it also becomes enlightened. 
Additionally, if a ${\sf profess}$ message is received from a processor with a
higher priority 
(as determined by the lexicographic comparison in line~\algoref{alg:lexi})
the processor sets $\ell$ to $0$.

{\emph{Compute step:}}
Processor $i$ updates its knowledge in $R_i[~]$ by including 
the information gathered from the received messages.
If processor $i$ receives a message $m$ such that 
$m.\ell \geq \lceil \log{n} \rceil$, then it calls the
 {Estimation()} procedure to compute the needed probability estimates and halts.
Otherwise processor $i$ increments $r$ and moves
  to the next round.

\smallskip

\noindent
{\bf Estimation() subroutine:}
The subroutine, given in Figure~\ref{fig:subalg}, calculates an estimate $\tilde{p_j}$
of probability $p_j$ for every processor $j \in P$ and stores the result
in $Estimate[j]$. 
For a processor $j$ whose crash is detected
(due to the lack of a response),
we set $Estimate[j] = -1$.
In the next section we discuss the rationale behind the estimation computation
and the choice of  parameters $\Gamma$ and $\Gamma_1$.
The estimate $\tilde{p_j}$ is calculated as follows.
First the tuples in $R[j]$ are sorted according to the round number, then
the sum of the first $N$ result correctness indicators (recall that 1 means correct, 0 means incorrect)
is computed for the largest $N$ such that the sum remains inferior to $\Gamma_1$.
The estimate  $\tilde{p_j}$ is then computed as $\frac{\Gamma_{1}}{N}$.

\begin{figure}[t]
\sglspc{0.91}
\hrule \smallskip
\begin{algorithmic}[1]
{\small
\STATENO{\bf subroutine} Estimation$(R[1..n],Estimate[1..n])$
\STATEIND\T Let $\Gamma=  (4\lambda \log{(\frac{2}{\delta})})/{\epsilon^2}$ and let
$\Gamma_1 = 1 + (1 + \epsilon) \Gamma$
\STATEIND\T {\bf for each}  $j \in P$ {\bf do} 
\STATEIND\TT {\bf if} $\exists \tup{res,src,rnd} \in R[j] : res=-1$ {\bf then} 
\STATEIND\TTT $Estimate[j] \gets -1$
\STATEIND\TT {\bf else}
\STATEIND\TTT Let $S$ be the list of tuples $\tup{res,src,rnd}$ in $R[j]$,
\STATENO\TTT sorted by
the round number $rnd$ in ascending order
\STATEIND\TTT Let $N$ be s.t. $\sum_{k=1}^{N}S[k].res < \Gamma_1 \leq \sum_{k=1}^{N+1}S[k].res$
\STATEIND\TTT $Estimate[j] \gets \Gamma_{1}/{N}$
\smallskip
\hrule
}
\end{algorithmic}\BB
\caption{\rm Estimation of the probabilities for each $j\in P$.}\BB
\label{fig:subalg}
\end{figure}


\Section{Estimation of Processor Reliability}\label{estimation}
Getting an $(\varepsilon, \delta)$-approximation $\tilde{p_i}$ for $p_i$, 
for any   $\varepsilon, \delta >0$, 
where ${\bf Pr}[p_i(1-\varepsilon) \leq \tilde{p}_i \leq p_i(1+\varepsilon)] > 1 - \delta$,    
   might sound like a straight forward problem solvable 
 by collecting a \emph{sufficient} number of samples and selecting the majority as the outcome.
  However, such a solution is programmable 
   if we know the required number of samples \mbox{\emph{a priori}}. 
 In fact this number  will be 
    dependent on the values of $p_i$, $\varepsilon$ and $\delta$. 
Since the value of $p_i$ is unknown,
   we want the algorithm to terminate as early as possible,  once
   the useful computations are done, 
without reliance on   the value of $p_i$ as either an input or a bound. The algorithm should be able to 
   detect if sufficient number of samples are collected on the fly 
 to arrive at an $(\varepsilon, \delta)$-approximation. Below we explain this with 
 an example.

Suppose we have a random variable $X$, where $X \in \{0,1\}$,
such that ${\bf Pr}[X=0] = p$ and  ${\bf Pr}[X=1] = 1-p =q$. 
Consider the independent and identically distributed (iid) 
random variables $X_1, X_2, \cdots, X_m$ whose distribution is
that of $X$. 
Therefore, $ \mathbb{E}[X] = \mathbb{E}[X_1] =$$ \ldots$$=  \mathbb{E}[X_m] = q$.
Suppose we want to use the unbiased estimator 
$\frac{S_m}{m}$ of $q$, 
where $S_m =
\sum_{i=1}^{m}X_i$.
\emph {An estimator $T(X_1, X_2,$$\ldots$$, X_m) $ of a
parameter $\theta$ is called unbiased estimator of $\theta$ if 
$\mathbb{E}_{\theta}[T(X_1, X_2,$$\ldots$$, X_m)] = \theta$}~\cite{CB2001}. 
Let us choose $m = c\log{n}$, for some $c > 0$, 
in an attempt to have a reasonable number of trials. 

By a simple application Chernoff bounds 
we can show that for $\delta > 0$
$$
{\bf Pr}\left[\frac{S_m}{m} \geq (1+\delta) q\right]
 \leq e^{-\frac{mq\delta^2}{3}}
 \leq e^{-\frac{\delta^2cq\log{n}}{3}}
 \leq n^{-\frac{cq\delta^2}{3}}
$$
 
A similar relation can be shown for the case where
${\bf Pr}[\frac{S_m}{m} \leq (1-\delta) q] \leq n^{-\frac{cq\delta^2}{2}}$.
Observe that unless we have some prior information about 
the value of $q$ (or $p$), other than the trivial bound
 $0\leq q \leq 1$, we may not know what $c$ to choose
to determine the number of repetitions for
obtaining the desired accuracy for the estimation of $q$. 
Thus it is desirable to have an algorithm that has 
an online rule for stopping the computation. 

Subroutine Estimation()  in Figure~\ref{fig:subalg} is
  used for calculating an $(\varepsilon, \delta)$-approximation of $p_i$
as described above.
 Now we elaborate on the technical aspects of $(\varepsilon,
\delta)$-approximation and determine the value of $\delta$ for
our analysis to hold \whp{}.
For every processor $i \in P-F$  we further 
  bound the number of test tasks required to 
 compute $\tilde{p_i}$. 

 The idea behind the subroutine Estimation() is based
on the Stopping Rule Algorithm (\SRA{})
of Dagum {\em et al.}~\cite{DKLR1995}. 
For completeness we reproduce in Figure~\ref{fig:SRA}
this well-known algorithm 
for estimating the mean of a random variable with support in $[0,1]$, 
with  $(\varepsilon, \delta)$-approximation.
Let $Z$ be a random variable distributed in the interval $[0,1]$ with 
mean $\mu_Z$. Let $Z_1, Z_2, \ldots$ be independently and identically 
distributed according to $Z$ variables. 
We say the estimate $\tilde{\mu}_Z$ is  an $(\varepsilon, \delta)$-approximation of $\mu_Z$ if 
$ {\bf Pr}[\mu_Z(1-\varepsilon) \leq \tilde{\mu}_Z \leq \mu_Z(1+\varepsilon)]
 > 1 - \delta$ .

\begin{figure}[t!]
\hrule \smallskip
\begin{algorithmic}[1]
{\small
\STATENO {\bf input parameters:} $(\varepsilon, \delta)$ with $0 < \varepsilon < 1$, $\delta >0$
\STATE Let $\Gamma  = 4 \lambda \log{(\frac{2}{\delta})}/\varepsilon^2$ \TTT \co{$\lambda = (e -2) \approx 0.72$}
\STATE Let $\Gamma_1 = 1 + (1 + \varepsilon) \Gamma$
\STATE\algolabel{stop:line:two} {\bf initialize} $N \leftarrow 0, S \leftarrow 0$ 
\STATE\algolabel{stop:line:three}  {\bf while} $S < \Gamma_1$ {\bf do} $N \leftarrow N +1$; $ S \leftarrow S + Z_N$ 
\STATE\algolabel{stop:line:four} {\bf output:} $\tilde{\mu}_Z \leftarrow \frac{\Gamma_1}{N}$
\smallskip
\hrule
}
\end{algorithmic}\BB
\caption{\rm The Stopping Rule Algorithm (\SRA{}) for estimating $\mu_Z$.}
\label{fig:SRA}
\end{figure}

Let us define $\lambda = ( e -2) \approx 0.72$ and 
$\Gamma  = 4 \lambda \log{(\frac{2}{\delta})}/\varepsilon^2$.
Now, Theorem~\ref{stopping} (slightly modified, from~\cite{DKLR1995})
tells us that \SRA{} provides us with 
an  $(\varepsilon, \delta)$-approximation with the number 
of trials within 
$\frac{\Gamma_1}{\mu_Z}$ 
\whp{}, where $\Gamma_1 = 1 + (1 + \varepsilon) \Gamma$.

\begin{theorem}[{\it Stopping Rule Theorem}]\label{stopping}
Let $Z$ be a random variable in $[0,1]$ with $\mu_Z = \mathbb{E}[Z] >0$.
Let $\tilde{\mu}_Z$ be the estimate produced  and 
let $N_Z$ be the number of experiments that
\SRA{}
runs with respect 
to $Z$ on inputs $\varepsilon$ and~$\delta$. Then,\\ 
\TT$(i)$~${\bf Pr}[\mu_Z(1-\varepsilon) \leq \tilde{\mu}_Z \leq \mu_Z(1+\varepsilon)]
 > 1 - \delta$,\\
\TT$(ii)$ $\mathbb{E}[N_Z] \leq \frac{\Gamma_1}{\mu_Z}$,
and\\
\TT$(iii)$~${\bf Pr}[N_Z > (1+\varepsilon)\frac{\Gamma_1}{\mu_Z}] \leq \frac{\delta}{2}$
.
\end{theorem}

\SRA{} computes an $(\varepsilon, \delta)$-approximation 
with an optimal number of samplings, within a constant factor~\cite{DKLR1995},
thus \SRA{}-based method provides substantial
  computational savings.

First, we want to show that 
${\bf Pr}[N_Z
 > (1 + \frac{1}{\varepsilon})^2c \log{n}] \leq \frac{1}{n^\alpha}$ for some
$c > 0$ and $\alpha > 0$. 
Let us choose a $\delta =  \frac{2}{n^{\alpha}}$, 
for some $\alpha> 0$, then for any 
$\varepsilon > 0$ and $\Gamma_1  = 1 + (1 + \varepsilon) \Gamma$ we have 
\begin{displaymath}
\Gamma = 4\lambda \log{\left(\frac{2}{2/n^{\alpha}} \right)}/\varepsilon^2 
= 4\lambda \log{(n^{\alpha})}/\varepsilon^2 
   =  \frac{4\lambda\alpha \log{n}}{\varepsilon^2}.
\end{displaymath}

\noindent
Also, we have 
$\Gamma_1 \leq (1+ \varepsilon) \frac{4\lambda\alpha'\log{n}}{\varepsilon^2}$ for some
$\alpha' > \alpha$. Now, using the Stopping Rule Theorem (Theorem~\ref{stopping}) 
we have 
\begin{eqnarray*} \label{ineq:one}
\frac{1}{n^{\alpha}}
\geq 
{\bf Pr}[N_Z > (1 + \varepsilon) \frac{\Gamma_1}{p_i}] 
  \geq 
{\bf Pr}[N_Z > (1 + \varepsilon)^2 \frac{4\lambda\alpha'
  \log{n}}{p_i\varepsilon^2}] = \\
 = 
{\bf Pr}[N_Z > (1+\frac{1}{\varepsilon})^2 \frac{4\lambda\alpha' \log{n}}{p_i}] 
   =  
{\bf Pr}[N_Z > (1+\frac{1}{\varepsilon})^2c \log{n}]
\end{eqnarray*}
where $c =  \frac{4\lambda\alpha'}{p_i} > 0$, i.e. $c = O(1)$.
Since we are interested in \whp{} guarantee, for a sufficiently large $n$, we can 
suitably  choose the constant $\alpha$, such that $\delta = \frac{1}{n^\alpha}$.

Our subroutine Estimation() is directly based on \SRA{}. 
To estimate  $p_i$ for $i \in P$ we need the sampling results (i.e., 
  the results of the test tasks). 
We compute the 
$(\varepsilon, \delta)$-approximation by looking at the history of the results 
stored in the list $S$ sorted in ascending order of the rounds
to consider the results in the order they where sampled.
Note that the results $Estimate_i[\;]$ may not be the same across all processors 
because the samples in $R_i[\;]$ may be different, however 
all we need is a sufficient number of results to compute an $(\varepsilon, \delta)$-approximation.

In our adaptation of \SRA{} to estimate $p_j$, in algorithm \AEalgo{}
   the corresponding random variable $Z$  
 takes the values $\{0,1\}$; $0$ for incorrect results and 
 $1$ for correct results. Note that in this case 
 we have a random variable $Z$, where $Z \in \{0,1\}$,
such that ${\bf Pr}[Z$$=$$1] = p_j$ and  ${\bf Pr}[Z$$=$$0] = 1-p_j =q_j$. 
Therefore, since $ \mathbb{E}[Z] = p_j$ we can estimate $p_j$ using  
 \SRA{}.
Based on the above derivation of a bound on $N_Z$ from 
 Theorem~\ref{stopping}
we know that, for every $p_i$, 
 $O(\log{n})$ computations of test task results, from processor $i$ are sufficient 
to  compute an
$(\varepsilon, \frac{1}{n^\alpha})$-approximation of ${p_i}$ by subroutine Estimation(), 
\whp{}. The following lemma summarizes this result. 

\begin{lemma}\label{lem:est}

In algorithm \Aest{}, subroutine Estimation() computes 
an $(\epsilon, \frac{1}{n^\alpha})$-approximation, for some constant $\alpha > 0$,
 of $p_i$ for any  $i \in P$, and the number of responses
from each live process $i$ sufficient for the estimation 
is  $O(\log{n})$, \whp{}.
\end{lemma}


\section{Complexity Analysis}\label{analysis}
Here we analyze the performance of algorithm \Aest{}.
We start by stating the Chernoff bound result, as well as, some lemmas and definitions
used in the analyses of our algorithm.

 \begin{lemma}[{\em Chernoff Bounds}]\label{chernoff}
 Let $X_1,X_2, \cdots, X_n$ be $n$ independent
 Bernoulli random variables with $\act{\bf Pr}[X_i=1] = p_i$ and
 $\act{\bf Pr}[X_i=0] = 1-p_i$,
 then it holds for $X=\sum_{i=1}^{n}X_i$
 and $\mu = \mathbb{E}[X] = \sum_{i=1}^{n}p_i$ that
 for all $\delta >0$,
 (i)~$\act{\bf Pr}[X \geq (1+\delta)\mu] \leq e^{-\frac{\mu\delta^2}{3}}$,
 and
 (ii)~$\act{\bf Pr}[X \leq (1-\delta)\mu] \leq e^{-\frac{\mu\delta^2}{2}}$.
 \end{lemma}

\begin{definition} [{\em The Coupon Collector's Problem (CCP)~\cite{RAJEEV95}.}]
\label{def:ccp} 
There are $n$ types of coupons and at each trial a coupon is chosen at random.
Each random coupon is equally likely to be of any of the $n$ types, and the random
choices of the coupons are mutually independent. Let $m$ be the number of trials.
The goal is to study the relationship between $m$ and the probability of having
collected at least one copy of each of $n$ types.
\end{definition}

In~\cite{RAJEEV95} it is shown that $\mathbb{E}[X]=n \ln{n} + O(n)$ and that \whp{} 
the number of trials for collecting all $n$ coupon
types lies in a small interval centered about its expected value.

Fraigniaud and  Glakkoupis~\cite{FG2010} 
study the communication complexity of rumor-spreading in the random
 phone-call model. They consider $n$ players communicating in parallel rounds, 
where in each round every player $u$ calls a randomly selected communication partner. 
Player $u$ is allowed to exchange information with the 
partner, either by {\em pulling} or {\em pushing} information.

The following lemma, proved in~\cite{FG2010}, shows that during the
{\em push} stage of the algorithm every rumor $\rho$ is 
disseminated to at least $\frac{3}{4}n$ players \whp{}.

\begin{lemma}~\cite{FG2010} \label{lemma-FG2010}
With probability $1- n^{-3+o(1)}$, at least $\frac{3}{4}$ fraction of the players
knows $\rho$ at the end of round $\tau = \lg{n} + 3 \lg{\lg{n}}$.
\end{lemma}

We next show that if a processor becomes \ENed{} then
every live processor terminates quickly.

\begin{lemma} \label{lemma:3}
In any execution of algorithm~\Aest{} ,
if a processor $q \in P-F$ is \ENed{} in round $\rho$,
then after additional $\Theta(\log{n})$ rounds every live processor
terminates \whp{}. 
\end{lemma}

\begin{Proof}
According to the  {\sc gossip} stage of the algorithm
if processor $q$ is \ENed{} then it starts sending ${\sf
  profess}$ messages. 
Without loss of generality we  assume that
$q$ is the processor with the highest priority among all \ENed{}
processors. According to  {\em Compute} step of {\sc gossip} stage
(line~\algoref{alg:stop} of algorithm~\Aest{}) every processor halts once it  receives a ${\sf
  profess}$ message $m$ from some processor such that $m.l \geq \lceil
\log{n} \rceil$. Since processor $q$
has the highest priority, once enlightened, it does not reset
its $\ell$ to $0$, and hence in $\Theta(\log{n})$ rounds of the
algorithm processor $q$ sends $\tilde{n} =cn \log{n}$ ${\sf profess}$ messages, where $c \geq
1$ is a constant. Let $r$ be the round in which processor $q$ sends
$\tilde{n}$ ${\sf profess}$ messages. 

We want to prove that in round $r$ every processor  receives a ${\sf
  profess}$ message from $q$ \whp{}. Let us assume that there exists
a processor $w$ that does not receive a ${\sf profess}$ message from
processor $q$ in round $r$. 
We prove that \whp{} such a processor does not exist.
Since $\tilde{n}$ ${\sf profess}$  messages are sent in round $r$,
there were $\tilde{n}$ random selections of processors from set $P$
in line~\algoref{alg:proc} by processor $q$;
let $i$ be the index of one such selection.
Let $X_i$ be a Bernoulli random variable such that $X_i = 1$ if
processor $w$ was chosen by processor $q$ and
$X_i=0$ otherwise. 

We define the random variable $X= \sum_{i=1}^{\tilde{n}}X_i$ 
to estimate the total number of times processor~$w$ is
selected in round $r$. 
In line~\algoref{alg:proc}
processor $q$ chooses a destination for the ${\sf profess}$  message
uniformly at random, and hence 
${\bf Pr}[X_i=1] = \frac{1}{n}$. 
Let $\mu = \mathbb{E}[X] = \sum_{i=1}^{\tilde{n}}X_i = \frac{1}{n}c
\, n \log{n} = c \log{n}$, then by applying Chernoff bound, for some
$1> \delta > 0 $, we have:

\begin{displaymath}
{\bf Pr}[ X \leq (1-\delta)\mu ] \leq e^{ -\frac{\mu\delta^2}{2}} 
\leq e^{-\frac{(c\log{n})\delta^2}{2}} \leq \frac{1}{n^{\frac{b\delta^2}{2}}} \leq \frac{1}{n^{\beta}}
\end{displaymath}
\noindent
where $\beta >0$. Hence,  ${\bf Pr}[ X \leq 1 ]  \leq \frac{1}{n^{\log{n}}}$.
Let $\mathcal{E}_w$ denote the fact that 
processor $w$ receives a message from processor $q$ in round $r$, and 
let $\bar{\mathcal{E}_w}$ be the complement of that event.  
By Boole's inequality we have 
${\bf Pr}[ \cup_w \bar{\mathcal{E}}_w ] \leq  \sum_w{\bf Pr}[\bar{\mathcal{E}_w} ] \leq \frac{1}{n^{\gamma}}$,
where $\gamma =\log{n} - 1 > 0$. Hence each processor $w \in
P-F$ receives at least one ${\sf profess}$  message from processor $q$ in
round $r$ \whp{}, i.e., 
${\bf Pr}[ \cap_w \mathcal{E}_w] = 
{\bf Pr}[ \overline{\cap_w\bar{\mathcal{E}_w}} ] 
 = 1- {\bf Pr}[ \cap_w\bar{\mathcal{E}_w} ] \geq  1 -
 \frac{1}{n^{\gamma}}$. Therefore, given that in round $r$ we have
 $q.l > \lceil \log{n} \rceil$, every live processor terminates in
 $\Theta(\log{n})$ rounds of the algorithm \whp{}.
\end{Proof}

Next lemma shows that if a processor $q \in P-F$ is \ENed{}, then in
each subsequent round 
$O(n\log{n})$ ${\sf profess}$  messages are sent
\whp{}.

\begin{lemma}\label{lem:term}
In the  {\em Send} step of {\sc gossip} stage of algorithm~\Aest{} 
$O(n\log{n})$ ${\sf profess}$ 
messages are sent in every round \whp{}.
\end{lemma}

\begin{Proof}
We use induction on the round number, by
showing that in every round there can be at most $kn\log{n}$ messages
for a sufficiently large constant such that $k>8$. Unless stated
otherwise, hereafter by messages we mean 
messages of  ${\sf profess}$  type that are being sent in the {\em Send} step
of {\sc gossip} stage.

The base case is the first round, say round $t_0$,  in which some set of  processors  sets 
  their {\it enlightened} variable to ${\sf true}$. 
 There can be at most  $n$  such processors, and
according to our algorithm, after {\it enlightened} is set to ${\sf true}$ for a
processor, it starts with $\ell=0$ and  sends
$\frac{1}{2}\log{n}$  ${\sf profess}$ messages, and hence, 
$O(n\log{n})$ ${\sf profess}$
messages are sent during round $t_0$. 
Let $M_{t}$ be  the set of messages 
sent by all processors in round $t$. 
Note that in round $t_0$ we have  $|M_{t_o}|\equiv m_{t_o} \leq k n \log{n}$. 

\noindent
{\em Induction hypothesis}: 
In round $t > t_0$ we have 
  $m_t  \leq  k n \log{n}$.

\noindent
{\em Induction step}: We want to show that in round $t+1$ we have
$m_{t+1} \leq kn \log{n}$.

Consider the processors 
at the beginning of round $t+1$. Observe that any message
from a processor with a higher priority to the processor with a lower
priority will reset $\ell=0$ at the latter processor.

Let $d_i$ denote the number of messages sent by the processor $i \in
P$ in  {\em Send} step of {\sc gossip} stage of round $t$. 
By the construction of the algorithm $d_i= 2^{\ell-1} \lceil
\log{n}\rceil$  where $\ell$ is the level of a processor and  $ 0 \leq
\ell  \leq \lceil \log{n} \rceil $. Note that 
any two distinct processors $i,j \in P$
can be at different levels ($\ell_i \neq \ell_j$).
Let us assume that the processor $id$'s are ranked in the descending
 order of the $d_i$'s. Hereafter when we refer to the $i$'th processor
 we mean the processor with ranking $i$, based on  $d_i$.

We define a random variable $X_i^{t}$ for each processor $i \in P$.
After all messages are sent and received in round $t$
we let $X_i^{t} = 0$  if processor $i$ received a message from  a
processor $j$ with a higher priority, and $X_i^{t}=1$ otherwise.
Let us further denote by $p_i ={\bf Pr}(X_i^{t}=1)$, note that 
$p_0=1$ 
since the processor $0$ has the highest priority.
Therefore, $p_0 = 1$;
$p_1 = (1-\frac{1}{n})^{d_0}$; $p_2 = (1-\frac{1}{n})^{d_0+d_1}$;
 $\ldots$ $p_i =(1-\frac{1}{n})^{\sum_{j=0}^{i-1}d_i}$.

We define ${\mathbf X}^{t}  = \sum_{i=0}^{n-1}d_i X_i^{t}$ 
as a random variable that
counts the number of messages that are sent 
during round $t+1$. Clearly, ${\mathbf X}^{t+1} \leq 2
  \sum_{i=0}^{n-1}d_i X_i^{t} + n \log{n}$.
The expected number of messages sent in round $t+1$ is bounded by:
$$
2\mathbb{E}\sum_{i=0}^{n-1}d_iX_i^{t}  + n\log{n} = 
2\sum_{i=0}^{n-1}d_i \mathbb{E}[X_i^{t}] + n\log{n} 
 = 2(d_0+\sum_{i=1}^{n-1}d_ic^{\sum_{j=0}^{i-1}d_j}) + n\log{n}
$$
\noindent
where $c \equiv c(n)= 1$$-$$\frac{1}{n}$. 
Consider the descending arrangement 
of $d_i$'s grouped in blocks of consecutive terms as 
$$
\underbrace{d_0, d_1 \cdots d_{k_1-1}}, 
 \underbrace{d_{k_1} \cdots d_{k_2-1}},  
\cdots 
\underbrace{d_{k_s } \cdots d_{n-1}}
$$
\noindent
where each group includes a maximum number of $d_i$'s such that $\sum_{i=k_j 
    }^{k_{j+1}-1}d_i <  n\log{n} $,  with a possible exception for the
    last block, where $j=0,1, \cdots, s$, $k_0 = 0$, and $k_{s+1}=n$.
We note that at the minimum the first grouping of $d_i$'s is within
the constant factor of $n \log{n}$, otherwise the total number of
messages sent is less than $k n \log{n}$ and the inductive step holds
for round $t+1$.
Using such blocking and
the fact that  $c < 1$ and $d_i \geq 0$ we have
%
\begin{eqnarray*}
\lefteqn{\sum_{i=1}^{n-1}d_ic^{\sum_{j=0}^{i-1}d_j} \leq }\\ 
&& \leq  \sum_{i=1}^{k_1-1}d_i  +   
\sum_{i=k_1 }^{k_2-1}d_ic^{\sum_{j=0}^{i-1}d_j} +  \cdots +
\sum_{i=k_s }^{n-1}d_ic^{\sum_{j=0}^{i-1}d_j} \\
&& \leq  
\sum_{i=1}^{k_1-1}d_i  +   
\sum_{i=k_1 }^{k_2-1}d_ic^{\sum_{j=0}^{k_1-1}d_j} +  \cdots +
\sum_{i=k_s}^{n-1}d_ic^{\sum_{j=0}^{k_s-1}d_j} \\
&& \leq
\sum_{i=1}^{k_1-1}d_i  +   
\sum_{i=k_1}^{k_2-1}d_i\frac{1}{n} +  \cdots +
\sum_{i=k_s }^{n-1}d_i\frac{1}{n^{s-2}}  
\leq  2 n \log{n}
\end{eqnarray*}

\remove{
\noindent
$
{\sum_{i=1}^{n-1}d_ic^{\sum_{j=0}^{i-1}d_j} \leq }$

$ \leq  \sum_{i=1}^{k_1-1}d_i  +   
\sum_{i=k_1 }^{k_2-1}d_ic^{\sum_{j=0}^{i-1}d_j} +  \cdots +
\sum_{i=k_s }^{n-1}d_ic^{\sum_{j=0}^{i-1}d_j} $

$ \leq  
\sum_{i=1}^{k_1-1}d_i  +   
\sum_{i=k_1 }^{k_2-1}d_ic^{\sum_{j=0}^{k_1-1}d_j}$$+$$ \cdots$$+
\sum_{i=k_s}^{n-1}d_ic^{\sum_{j=0}^{k_s-1}d_j} $

$ \leq
\sum_{i=1}^{k_1-1}d_i  +   
\sum_{i=k_1}^{k_2-1}d_i\frac{1}{n} +  \cdots +
\sum_{i=k_s }^{n-1}d_i\frac{1}{n^{s-2}}  
\leq  2 n \log{n}
$
}

\smallskip\noindent
since  $c(n)^{n\log{n}} \rightarrow \frac{1}{n}$ as $n \rightarrow \infty$.
Therefore, we have 
$$\mathbb{E}[X^{t+1}] \leq 2\mathbb{E}\sum_{i=0}^{n-1}d_iX_i^{t}  +
n\log{n} \leq 7 n\log{n} $$

By Chernoff bound with negative dependencies for some $\delta > 0$ we
have
$${\bf Pr}( {\mathbf X}^{t+1} \geq (1+\delta)\mathbb{E}[ {\mathbf
 X}^{t+1}]) \leq  e^{-\frac{1}{2}\mathbb{E}[ {\mathbf X}^{t+1}]\delta^2}  
\leq e^{-\frac{1}{2}7n \log{n}\delta^2}  \leq  \frac{1}{n^\beta}$$

\noindent
where $\beta$ is some positive constant.  
\end{Proof}

To simplify the presentation we proceed by defining the {\em
  estimability} property, that tells us whether enough samples have
been gathered.

\begin{definition}{\it (Estimability)}
\label{def:estim}
We say that probability $p_j$ is \emph{estimable} for 
$j \in P$ in round $r$ of algorithm \Aest{}, 
if at the  end of round $r$ we have 
$\sum_{ x \in \bigcup_{i\in P-F }R_{i}[j] }I_{\{1\}}x.res \geq
\Gamma_1$, or for some processor $i \in P-F$, $\exists x\in R_i[j]$ such that $x.res=-1$.
\end{definition}

In the previous section we  showed
that the number of responses sufficient to estimate $p_i$ with
$(\epsilon, \frac{1}{n^\alpha})$-approximation using subroutine Estimation() is
$O(\log{n})$. (In the sequel we let $\delta$ stand for $
\frac{1}{n^\alpha}$.)
We next assess the number of rounds required for a processor $i \in P-F$ to
become \ENed{}, that is the number of rounds required for 
$i$ either to collect sufficient responses for every processor $j \in P$ or
to possess the result $-1$ from $j$, indicating that it crashed.
The analysis follows along the lines of the
analysis done in our earlier papers~\cite{DKRS2013, DKS2011};
except that here we argue about random selection of processors versus tasks in our
prior work.
Due to paucity of space  we refer the kind reader to 
\cite{DKRS2013, DKS2011} when appropriate to avoid a restatement of
our results.

{In \emph{Compute} step of  {\sc query} stage
a processor does at most $\lceil \log{n} \rceil$ tasks.
Thus, it is possible that a live processor will not respond to a
request to perform a test task. In this aspect the algorithm differs from
the approach in~\cite{DKRS2013, DKS2011} where if a task is selected by a
live processor, then it is consequently executed. Fact~\ref{fact:thr}
 below (a rewording after~\cite{ABKU1999}) shows that
\whp{} no processor receives more than $\lceil \log{n}
\rceil$ requests in one round.
\begin{fact}\label{fact:thr}
If  $n$ balls are uniformly randomly placed into $n$ bins with probability 
at least $1 - \frac{1}{n^c}$, for some $c>0$, the fullest bin has 
$(1 + o(1))\frac{\log{n}}{\log{\log{n}}}$ balls.
\end{fact}

We now analyze our algorithm in the three adversarial models.
Let $F_r$ be the set of processors crashed before round~$r$.

\subsection{Analysis of Algorithm~\AEalgo{} for Failure Model  \M{\ell f}} \label{model:f}
Here $|F_r|$ is bounded as in model \M{} of~\cite{DKS2011} with 
at most $hn$ processor crashes for a constant $h \in (0,1)$.
Next lemma determines the number of rounds required for
algorithm~\AEalgo{} in model \M{\ell f} so that \whp{} $p_j$ is {\em estimable} 
for every processor  $j \in P$.

\begin{lemma} \label{lem:complF}
In any execution of algorithm~\AEalgo{} under the failure model
\M{\ell f}, after $O(\log{n})$ rounds  $p_j$ is {\em estimable}
for every processor  $j \in P$, \whp{}. 
\end{lemma}

\begin{Proof}
According to Lemma~\ref{lem:est} the number of responses from each
live processor $i$ sufficient for subroutine Estimation() to compute an
$(\varepsilon, \frac{1}{n^\alpha})$-approximation of $p_i$ is
$O(\log{n})$ \whp{}. Let $\tilde{n}=k\log{n}$ be the number of
responses sufficient to estimate $p_i$ for any processor $i \in P$, where
 $k>0$ is a sufficiently large constant.
From above, and from the definition of {\em
  estimability}, it follows that the probability $p_j$ is estimable for a
live processor $j \in P-F$ at the end of some round $r$ if processors in $P-F$
collectively possess $\tilde{n}$ results from processor $j$.
 On the other hand, if a processor $j$ crashes prior to the round $r$
 then $p_j$ is {\em estimable} if either by round $r$ it executed at least
 $\tilde{n}$ tasks assigned to it by processor in $P-F$, or a
 processor $i \in P-F$ did not recieve a response from $j$
 (line~\algoref{crash} of algorithm \AEalgo{}), after
 sending a task to $j$ 
 (lines~\algoref{proc:select}-\algoref{send:task} of
 algorithm \AEalgo{}). 

We want to show that \whp{} after
$r=\kappa \tilde{n}$ rounds of algorithm \AEalgo{}, where $\kappa = \frac{1}{1-f}k$ is a
constant, every live processor $j \in P-F$ executes at least
$\tilde{n}$ tasks assigned to it by processors in $P-F$.
Conversely, based on the Fact~\ref{fact:thr}, we want to show that every processor  
$w \in P$ is selected by  processors in
$P-F$ to execute a task at least $\tilde{n}$ times by round $r$.  Note
that, in the latter case, by the argument provided above, it follows
that $p_w$ is estimable for every processor $w \in P$, whether live or not.

Let us assume that after $r= \kappa \tilde{n}$ rounds of algorithm
\AEalgo{} there exists a processor $w \in P$, such that it is
selected by processors in $P-F$ to execute a task less than
$(1-\delta_1)k\log{n}$ times, for some $\delta_1 > 0$. We
prove that \whp{} such a processor does not exist.

\sloppy{According to our assumption at the end of round $r$ for some
processor $w$, we have }
$|\bigcup_{i\in P-F }R_{i}[w]|  < (1-\delta_1)k \log{n}$.
We prove that for any processor $w \in  P$ \whp{} the latter cannot
happen. This is because even if $w$ crashes prior to some round $r'
<r$ and a processor $i \in P-F$
assignes a task to $w$ in round $r'$ then $\langle -1,i,r' \rangle$ is
added to $R_i[w]$ according to line~\algoref{crash} of algorithm
\AEalgo{}.

Let $X_i$ be a Bernoulli random variable such that $X_i = 1$ if
processor $w$ was chosen to perform a task in line~\algoref{proc:select}
of the
algorithm by a processor in $P-F$, and
$X_i=0$ otherwise. Based on the adversarial model \M{\ell f}, we know that
$|P-F| \geq (1-f)n$, where $f \in (0,1)$.

Let us next define the random variable $X=X_1+ ... + X_{r(1-f)n}$ to
count the total number of times processor~$w$ is
selected by processors in $P-F$ by the end of $r$ rounds. 
Note that according to line~\algoref{proc:select}
any processor chooses a processor from $P$ for executing a test task
uniformly at random, and hence 
${\bf Pr}[X_i=1] = \frac{1}{n}$. 
Let 
$\mu = \mathbb{E}[X] = \sum_{i=1}^{r(1-f)n}X_i = \kappa \tilde{n} (1-f)n \cdot \frac{1}{n}= \frac{1}{1-f}k
\, (1-f)n \log{n} = k \log{n}$, then by applying the Chernoff bound, for the same
$\delta_1$ chosen as above, we have:
\begin{displaymath}
{\bf Pr}[ X \leq (1-\delta_1)\mu ] \leq e^{ -\frac{\mu\delta_1^2}{2}} 
\leq e^{-\frac{(k\log{n})\delta_1^2}{2}} \leq \frac{1}{n^{\frac{b\delta_1^2}{2}}} \leq \frac{1}{n^{\alpha'}}
\end{displaymath}
where $\alpha' >1$ for some sufficiently large $b$. 
Thus, we have ${\bf Pr}[ X \leq (1-\delta_1)k\log{n}] \leq
\frac{1}{n^{\alpha'}}$ for some $\alpha' > 1$.
Now let us denote by $\mathcal{E}_w$ the fact that 
 $|\bigcup_{i\in P-F }R_{i}[w]|  > (1-\delta_1)k \log{n}$ by the end of round $r$, and 
let $\bar{\mathcal{E}_w}$ be the complement of that event.  
By Boole's inequality we have 
${\bf Pr}[ \cup_w \bar{\mathcal{E}}_w ] \leq  \sum_w{\bf Pr}[\bar{\mathcal{E}_w} ] \leq \frac{1}{n^{\beta}}$,
where $\beta =\alpha' - 1 > 0$. Hence each processor $w \in P$
is the destination of at least $(1-\delta_1)k \log{n}$ test task
execution requests
\whp{}, i.e., 
$$
{\bf Pr}[ \cap_w \mathcal{E}_w] = 
{\bf Pr}[ \overline{\cup_w \bar{\mathcal{E}_w}} ] 
 = 1- {\bf Pr}[ \cup_w \bar{\mathcal{E}_w} ] 
\geq  1 -\frac{1}{n^{\beta}} \; .
$$

\noindent
Hence $p_w$ is {\em estimable} \whp{}. This completes the proof.

\end{Proof}

The proof of the next lemma is similar to the proof of Lemma~6
in~\cite{DKS2011}.

\begin{lemma} \label{lem:dissF}
In any execution of algorithm~\AEalgo{} under failure model \M{\ell f},
if $p_j$ is {\em estimable} in round $\rho$ for every  processor $j \in P$
then, after additional $O(\log{n})$ rounds, at least one processor from $P-F$ is \ENed{}
\whp{}. 
\end{lemma}

\begin{Proof}
Let us assume that in some round $r$ processor $i \in P-F$ selects
some processor $j \in P$ and assignes a test task $t$ to it. According to
algorithm \AEalgo{} a triple $ \vartheta \equiv \langle
v_j,i,r\rangle$ is added by processor $i$ to $R_i[j]$,  
where $v_j$ is $1$ if $t$ was computed correctly by $j$, $0$ if it was
computed incorrectly, and $-1$ if processor $j$ did not respond to
$i$. According to Fact~\ref{fact:thr} the latter means that $j$
crashed \whp{}. Based on Lemma~\ref{lem:complF} in $O(\log{n})$
rounds of algorithm \AEalgo{}, $p_j$ is estimable for every processor
$j \in P$, and hence, as we argued in the proof of
Lemma~\ref{lem:complF}, there are $O(\log{n})$ triples generated
for every processor $i \in P$. Let $\mathcal{V}$ be the corresponding set of
triples in the system. We want to prove that once a triple $\vartheta
\in \mathcal{V}$ is generated in the system by a processor in $P-F$ then \whp{} it takes
$O(\log{n})$ rounds for the rest of the processors in $P-F$ to learn
about $\vartheta$.

In model \M{\ell f} at most $fn$ processors may crash, where $f \in
(0,1)$. Thus, there are $\Theta(n)$ processors left in $P-F$. Hence,
we can apply Lemma~\ref{lemma-FG2010} to algorithm \AEalgo{} and we infer
that in $O(\log{n})$ rounds of the algorithm  at least $\frac{3}{4}n$ of processors
in $P-F$ become aware of triple $\vartheta$ \whp{}.
 Next consider any round $d$ such that at least $\frac{3}{4}n$ of the
 processors in $P-F$ are aware of triple 
$\vartheta$ for the first time. Let us denote this subset of processors by $S_d$ 
($|S_d| \geq \frac{3}{4}n$.)

We denote by $U_d$ the remaining fraction of the processors from $P-F$ that are not aware of
$\vartheta$. We are interested in the number of rounds 
required for every worker in $U_d$ to learn about $\vartheta$ \whp{} by receiving a 
message from one of the workers in $S_d$ in some round following $d$.

We show that, by the analysis very similar to the Coupon's Collector
Problem (Definition~\ref{def:ccp}),
 in $O(\log n)$ rounds triple $\vartheta$ is known to all
 processors in $P-F$ \whp{}.
Every processor in $P-F$ has a unique id, hence we can think of those processors
as of different types of coupons and we assume that the processors in $S_d$ collectively
represent the coupon collector. In this case, however, we do not require that
every processor in $S_d$ contacts all processors in $U_d$ \whp{}. 
Instead, we require only that
the processors in $S_d$ \emph{collectively} contact all processors in $U_d$ \whp{}.
According to our algorithm, if no processor from $P-F$ is \ENed{}, in
every round every processor in $P-F$ $(S_d \subset P-F)$, 
selects a processor uniformly at random and sends all its data 
to it in a $\act{share}$ message (line~\algoref{share} of the algorithm). Let us denote
by $m$ the collective number of trials by processors in $S_d$ to contact processors in $U_d$.
According to CCP if $m= O(n \ln{n})$ then \whp{} processors in $S_d$ collectively contact every 
processor in $P-F$, including those in $U_d$. Since there are at least $\frac{3}{4} (1-f)n$ 
processors in $S_d$ then in every round the number of trials is at least $\frac{3}{4} (1-f)n$,
hence in $O(\log{n})$ rounds \whp{} all processors in $U_d$ learn about
$\vartheta$. Note, that the number of rounds may increase by a
constant factor
of $\frac{1}{1-f}$ in comparison to the case when there are no
crashes, however this does not affect our asymptotic results.
Therefore, in $O(\log n)$ rounds \whp{} all processors in
$U_d$ learn about $\vartheta$.

Thus we showed that if a new triple is generated in the system then \whp{} it will
be known to all live processors in $O(\log n)$ rounds.
Now by applying Boole's inequality we want to show that \whp{} in $O(\log n)$ rounds  
all generated triples are spread among all live processors.

Since $|P|=n$ there are $O(n\log n)$ triples in $\mathcal{V}$ by the
time every processor in $P$ is estimable.
Let $\overline{\mathcal{E}}_\vartheta$ be the event that some triple
$\vartheta \in \mathcal{V}$ is not spread around
among all live processors in $\tilde{r} = k \log{n}$ rounds weher
$k>0$ is a sufficiently large constant. In the preceding part of the proof 
we have shown that ${\bf Pr}[\overline{\mathcal{E}}_\vartheta] < \frac{1}{n^\beta}$,
where $\beta>1$. By Boole's inequality, the probability that there exists one
 triple that did not get spread to all live workers, can be bounded as 
\begin{displaymath}
{\bf Pr} [\cup_{\vartheta \in {\mathcal
    V}}\overline{\mathcal{E}}_\vartheta ] \leq \Sigma_{\vartheta \in
  {\mathcal V}}{\bf Pr}[\overline{\mathcal{E}}_\vartheta] = 
O(n \log n)\frac{1}{n^\beta} \leq \frac{1}{n^\gamma}
\end{displaymath}
where $\gamma>0$. This implies that upon termination
every live processor collects all $O(n \log n)$
triples generated in the system \whp{}. Thus, at least one processor
in $P-F$ becomes \ENed{} after $O(\log{n})$ rounds \whp{}.
\end{Proof}

Next we assess time complexity, work complexity, 
and message complexity of algorithm~\AEalgo{} under the failure model
\M{\ell f}.

\begin{theorem}\label{compl:est:lf}
For every processor $i \in P-F$ algorithm~\AEalgo{} computes 
an $(\varepsilon, \delta)$-approximation of $p_i$,
for the given $\delta >0$ and $\varepsilon > 0$, under the failure
model \M{\ell f},
with time complexity $\Theta(\log{n})$, work complexity $\Theta(n\log{n})$, and
message complexity $\Theta(n\log^2n)$.
\end{theorem}

\begin{Proof}
According to Lemma~\ref{lemma:3} once a processor $q \in P-F$ is
\ENed{}, algorithm~\AEalgo{} terminates after additional $\Theta(\log{n})$ rounds
\whp{}. On the other hand, according to Lemmas~\ref{lem:complF}
and~\ref{lem:dissF} at least one processor from $P-F$ is \ENed{}
in $O(\log{n})$ rounds of algorithm~\AEalgo{}. Hence, the time
complexity of the algorithm is $\Theta(\log{n})$.
There are $\Omega(n)$ live processors in every round, and hence, the work complexity
of the algorithm is  $\Theta(n\log{n})$.

Lastly, according to Lemma~\ref{lem:term}, once a processor is
\ENed{},  in the {\em Send} step of the
{\sc gossip} stage $O(n \log{n})$ ${\sf profess}$ messages are sent in every round
\whp{}. Notice, that $O(n)$ messages are sent in every round if no
processor is \ENed{}. On the other hand, according to
Lemma~\ref{lemma:3}, once a processor from $P-F$ is enlightened,
algorithm~\AEalgo{} terminates after $\Theta(\log{n})$
rounds. Hence, the message complexity of
the algorithm is $\Theta(n\log^2n)$.
\end{Proof}

\subsection{Analysis of Algorithm~\AEalgo{} for Failure Model \M{fp}} \label{model:F-fp}
In model\M{fp} we have $|F| \le n - n^a$.
For the purpose of analysis we divide an execution
of the algorithm into two epochs:
\epoch{a} consists of all rounds $r$ where $|F_r|$ is at most linear in $n$, 
so that when the number of live processors is at least $c'n$
for some suitable constant $c'$;
\epoch{b} consists of all rounds $r$ starting with first round $r'$ 
(it can be round 1) when the number of live
processors drops below some $c'n$ and becomes
$c''n^a$ for some suitable constant $c''$.
Note that either epoch may be empty.

For the small number of crashes in \epoch{a}, 
Theorem~\ref{compl:est:lf} in Section~\ref{model:f} gives the worst case
work as $\Theta(n \log n)$ and message complexity as $\Theta(n \log^{n})$;
the upper bounds apply whether or not the algorithm
terminates in this epoch.

Next we consider \epoch{b}.
If the algorithm terminates in round $r'$, the first round of the epoch,
the cost remains the same as
given by Theorem~\ref{compl:est:lf}.
If it does not terminate, it incurs additional
costs associated with the processors in $P - F_{r'}$,
where $|P - F_{r'}| \le c'' n^a$. 
We analyze the costs for \epoch{b} in the rest
of this section. The final message and work complexities
will be at most the worst case complexity for \epoch{a}
plus the additional costs for \epoch{b}
incurred while $|P-F|= \Omega(n^a)$ per model \M{fp}.

\begin{lemma} \label{lem:complFp}
 In any execution of algorithm~\AEalgo{} under failure model \M{fp},
 after $O(n^{1-a}\log{n})$ rounds of \epoch{b}
 $p_j$ is {\em estimable} for every processor $j \in P$,  \whp{}. 
\end{lemma}

\begin{proofsketch}
The proof of the lemma is easily obtained by arguing along the lines
of Lemma~\ref{lem:complF} and Lemma~5 of~\cite{DKRS2013}.
\end{proofsketch}

\begin{lemma} \label{lem:dissFp}
In any execution of algorithm~\AEalgo{} under failure model
\M{fp}, 
if $p_j$ is {\em estimable} in round $\rho$ for every processor $j \in P$,  
then, after additional $O(n^{1-a} \log{n} \log{\log{n}})$ rounds of
\epoch{b}, at least one processor in $P-F$ is \ENed{}, \whp{}. 
\end{lemma}

\begin{proofsketch}
The proof of this lemma is easily obtained by arguing along the lines
of Lemma~\ref{lem:dissF} and Lemma~8 of~\cite{DKRS2013}.
\end{proofsketch}

\begin{theorem}\label{compl:est:fp}
For every processor $i \in P-F$ algorithm~\AEalgo{} computes
an $(\varepsilon, \delta)$-approximation of $p_i$,
for the given $\delta >0$ and $\varepsilon > 0$, under the failure
model \M{fp},
with time complexity $O(n^{1-a}\log{n}\log{\log{n}})$, work complexity $O(n\log{n}\log{\log{n}})$, and
message complexity $O(n\log^2n)$.
\end{theorem}

\begin{Proof}
To obtain the result  we combine the costs associated with 
\epoch{a} with the costs of \epoch{b}.
The work and message complexity bounds for \epoch{a} are given
by Theorem~\ref{compl:est:lf} 
and are  $\Theta(n \log n)$ and  $\Theta(n \log^2 n)$ respectively.

For \epoch{b} (if it is not empty), 
where $|P-F| = O(n^a)$,  per
Lemmas~\ref{lemma:3},~\ref{lem:complFp} and~\ref{lem:dissFp}
the algorithm terminates after $O(n^{1-a}\log{n}\log{\log{n}})$
rounds \whp{} and there are $\Theta(n^{a})$ live processors, thus its work is 
$O(n\log{n}\log{\log{n}})$. 

On the other hand, according to Lemma~\ref{lem:term}, once a processor is
\ENed{},  in the {\em Send} step of the
{\sc gossip} stage $O(n \log{n})$ ${\sf profess}$ messages are sent in every round
\whp{}. Notice, that in \epoch{b} $O(n^a)$ messages are sent in every round if no
processor is \ENed{}. On the other hand, according to
Lemma~\ref{lemma:3}, once a processor from $P-F$ is enlightened,
algorithm~\AEalgo{} terminates after $\Theta(\log{n})$
rounds. Hence, the message complexity of
the algorithm is $O(n\log^2n)$.

The worst case costs of the algorithm correspond to the executions
with non-empty \epoch{b}, where the algorithm does not terminate early. 
In this case the costs from  \epoch{a}
are asymptotically absorbed into the worst case costs of \epoch{b}
computed above.
\end{Proof}

\subsection{Analysis of Algorithm~\AEalgo{} for Failure Model \M{pl}} \label{model:F-pl}
{In the adversarial  model \M{pl} we have 
$|P-F| = \Omega(\log^c{n})$.}
For executions in \M{pl}, let $|P-F|$ be at least $b \log^c n$, 
for specific constants $b$ and $c$ satisfying the model constraints.
Let $F_r$ be the actual number of crashes that occur
prior to round $r$. 
For the purpose of analysis we divide an execution
of the algorithm into two epochs:
\epoch{b'} consists of all rounds $r$ where $|F_r|$ 
remains bounded as in model \M{fp}
(for reference, this epoch combines \epoch{a} and \epoch{b}
from the previous section);
\epoch{c} consists of all rounds $r$ starting with the first round $r''$ 
(it can be round 1) when the number of live
processors drops below  
$b' n^a$, where $b'$ and $a$ are specified by the failure model
 \M{fp}, but remains
  $\Omega(\log^c{n})$ per model \M{pl}. Observe that since we are concerned with model
\M{pl}, in the sequel we can
  chose any $a$, such that $0<a<1$.
Also note  that either epoch may be empty.

In \epoch{b'} the algorithm incurs costs exactly as
in model \M{fp}.
If  algorithm \AEalgo{} terminates in round $r''$, the first round of the epoch,
the costs remain the same as the costs analyzed for \M{fp}
in the previous section.

If it does not terminate, it incurs additional
costs associated with the processors in $P - F_{r''}$,
{where $b \log^c n \le |P - F_{r''}| \le b' n^a$.} We analyze the costs for \epoch{c} next. 
The final message and work complexities
are then at most the worst case complexity for \epoch{b'}
plus the additional costs for \epoch{c}.

In the next lemmas we use the fact that $|P - F_{r''}|=\Omega(\log^c{n})$.
The first lemma shows that within some $O(n)$ rounds in \epoch{c}
$p_j$ is {\em estimable} for every $j\in P$, \whp{}. 

\begin{lemma} \label{lem:complFl}
In any execution of algorithm~\AEalgo{} under failure model \M{pl},
after $O(n)$ rounds of   \epoch{c}
  $p_j$ is {\em estimable} for every $j\in P$,   \whp{}. 
\end{lemma}

\begin{proofsketch}
The proof of the lemma is easily obtained by arguing along the lines
of Lemma~\ref{lem:complF} and Lemma~9 of~\cite{DKRS2013}.
\end{proofsketch}

\begin{lemma} \label{lem:dissFl}
In any execution of algorithm~\AEalgo{} under failure model \M{pl},
if  $p_j$ is {\em estimable} in round $\rho$ for every $j\in P$,  
then,  after $O(n)$ rounds of \epoch{c}, at least one processor in $P-F$ is \ENed{}, \whp{}. 
\end{lemma}

\begin{proofsketch}
The proof of this lemma is easily obtained by arguing along the lines
of Lemma~\ref{lem:dissF} and Lemma~10 of~\cite{DKRS2013}.
\end{proofsketch}

Next we assess time complexity, work complexity, 
and message complexity of algorithm~\AEalgo{} under the failure model
\M{pl}. 

\begin{theorem}\label{compl:est:pl}
For every processor $i \in P-F$ algorithm~\AEalgo{} computes
an $(\varepsilon, \delta)$-approximation of $p_i$,
for the given $\delta >0$ and $\varepsilon > 0$, under the failure
model \M{pl},
with time complexity $O(n)$, work and message complexities $O(n^{1+a})$.
\end{theorem}

\begin{Proof}
To obtain the result  we combine the costs associated with 
\epoch{b'} with the costs of \epoch{c}.
As reasoned earlier, the worst case costs for \epoch{b'}
are given in Theorem~\ref{compl:est:fp}.

For \epoch{c} (if it is not empty), 
where $|P-F| = \Omega(\log^c{n})$, per
Lemmas~\ref{lemma:3},~\ref{lem:complFl} and~\ref{lem:dissFl},
algorithm \AEalgo{} terminates  
after $O(n)$ rounds \whp{} and 
there are up to $O(n^a)$ live processors, thus its work is 
$O(n^{1+a})$. 

On the other hand, according to Lemma~\ref{lem:term}, once a processor is
\ENed{},  in the {\em Send} step of the
{\sc gossip} stage $O(n \log{n})$ ${\sf profess}$ messages are sent in every round
\whp{}. Notice, that in \epoch{c}, when no processor is \ENed{},
$O(n^a)$ messages are sent in every round.  On the other hand, according to
Lemma~\ref{lemma:3}, once a processor from $P-F$ is enlightened,
algorithm~\AEalgo{} terminates after $\Theta(\log{n})$
rounds. Hence, the message complexity of
the algorithm is $O(n^{1+a})$, for any $ 0<a<1$.

The worst case costs of the algorithm correspond to executions
with a non-empty \epoch{c}, where the algorithm does not terminate early. 
In this case the costs from  \epoch{b'}
are asymptotically absorbed into the worst case costs of \epoch{c}
computed above.
\end{Proof}

\begin{observation}
We note that it should be possible to derive tighter bounds
on the complexity of the algorithm. 
This is because we only assume for \epoch{c} that the number of live processors
is bounded by the generous range
$b \log^c n \le |P - F_{r}| \le b' n^a$. 
In particular, if in \epoch{c} there are $\Theta(poly \log{n})$ live processors,
the work and message complexities become $O(n \, poly\log{n})$
as follows from the arguments along the lines of the proofs of
Theorem~\ref{compl:est:pl}. 
\end{observation}

\Section{Conclusion} \label{conclusion}
We presented a synchronous decentralized  algorithm
that assesses reliability of processors in the context of 
cooperative distributed computing. Specifically,
we estimate the probabilities of processors performing
their tasks correctly as an $(\epsilon, \delta)$-approximation.
Our randomized algorithm is also able to deal with processor crashes.
We established  time, work, and message complexity analyses that demonstrate
the efficiency of the algorithm with high probability guarantees. 
The analysis was performed in three different models that differ
in the extent of crashes occurring during its execution.
We note that when our algorithm is used as a precursor
to network supercomputing, its costs are completely
amortized if there is a polylog number of tasks per processor.

\renewcommand{\baselinestretch}{1.0}\normalsize

\bibliographystyle{plain}
\bibliography{byzantine}

\end{document}